\documentclass[twoside,11pt]{report}
\usepackage{graphicx}
\usepackage{amssymb, amsmath, amsxtra}
\usepackage{epsfig,subfigure}

\usepackage{latexsym}
\usepackage{bm}
\usepackage{wrapfig}

 \usepackage{ntheorem}
           \theoremstyle{plain}
                      {\theorembodyfont{\rmfamily}
                      \theoremseparator{.}
                       \newtheorem{example}{Example}[section]
           \newtheorem{theorem}{Theorem}[section]
           \newtheorem{proposition}{Proposition}[section]
           \theoremstyle{plain}
           \newtheorem{cor}{Corollary}[section]
           \theoremstyle{plain} \newtheorem{lemma}{Lemma}[section]
           \theoremstyle{plain}
           \newtheorem{defn}{Definition}[section]
            
           \theoremstyle{plain}
           \newtheorem{remark}{Remark}[section]}

\begin{document}

 \begin{center}
{\Large \bf Group Analysis of Non-autonomous Linear \\[1ex] Hamiltonians through Differential Galois Theory.}\\[3ex]

\large{David Bl\'azquez-Sanz}\\[1ex]
{\it Instituto de Matem\'aticas y sus Aplicaciones (IMA), \\ Universidad Sergio Arboleda, \\ Bogot\'a, Colombia.} 
\\[3ex]

 \large{Sergio A. Carrillo Torres}\\[1ex]
{\it Facultad de Ciencias, \\ Departamento de Matem\'aticas \\ Universidad Nacional de Colombia, \\
Bogot\'a, Colombia.}
 \end{center}

 \noindent
{\bf Abstract}\\
In this paper we introduce a notion of integrability in the non autonomous sense. For the
cases of $1+\frac{1}{2}$ degrees of freedom and quadratic homogeneous Hamiltonians of $2+\frac{1}{2}$ degrees of
freedom we prove that this notion is equivalent to the classical complete integrability of the system
in the extended phase space. For the case of  quadratic homogeneous Hamiltonians of $2+\frac{1}{2}$ degrees of
freedom we also give a reciprocal of the Morales-Ramis result. We classify those systems 
by terms of symplectic change of frames involving algebraic functions of time, and give their canonical forms. \\[2ex]


\section{Introduction and Main Results}

Differential Galois theory has been fruitful applied to the study of
integrability of Hamiltonian systems, see for instance \cite{MoralesLibro, MoralesRamis1, MoralesRamis2, MoralesRamisSimo,
MoralesSimoSimon,MoralesSimon,Casale}
and many others. In a recent paper \cite{Acosta}, this Morales-Ramis approach
is applied to the study of non-autonomous Hamiltonian systems. In order to do that, we extend 
a non-autonomous Hamiltonian system of $n+\frac{1}{2}$ degrees of freedom to an autonomous
system of $n+1$ degrees of freedom, and we apply Morales-Ramis techniques to this last one.
However, there is a number of open questions here. How do the integrability of the original and extended
systems relate? The variational equation itself is a linear non-autonomous Hamiltonian system, but its extension
is not linear anymore. The Morales-Ramis approach provide us necessary condition for the integrability. 
However, in the linear context. Do the Morales-Ramis approach give us sufficient conditions for the 
integrability? We arrive to the following result. 

\medskip

{\noindent\bf Theorem \ref{main}.}\emph{
Let $H\in \overline{\mathbb M(\Gamma)}[x_1,x_2,y_1,y_2]_2$ be a quadratic homogeneous non-autonomous Hamiltonian
of $2+\frac{1}{2}$ degrees of freedom, with coefficients meromorphic in $\overline{\Gamma}$ a ramified
covering of $\Gamma$. The following are equivalent:
\begin{enumerate}
\item[(1)] The associated extended autonomous system $\widehat{H}= H + h$ is completely
integrable by meromorphic functions in $\widehat{\Gamma} \times V \times \mathbb C_h$ for some
ramified covering $\widehat{\Gamma}$ of $\overline{\Gamma}$.
\item[(2)] $H$ is integrable in the non-autonomous sense  by meromorphic functions 
in $\widehat{\Gamma} \times V$ for some ramified covering $\widehat{\Gamma} \times$ of $\overline{\Gamma}$.
\item[(3)] $H$ is integrable in the non-autonomous sense by quadratic first
integrals $F_1,F_2\in \overline{\mathbb M(\Gamma)}[x_1,x_2,y_1,y_2]_2$.
\item[(4)] The connected component of the Galois group of $\vec X_{H}$ is a abelian.
\end{enumerate}}

\medskip

In general, the classification of integrable systems is, in general an interesting and difficult problem. 
See for instance \cite{Nakagawa}, for related results.
In this work, we arrive to a classification 
of all integrable quadratic (and non-linear in $t$) $2+\frac{1}{2}$
degrees of freedom Hamiltonians. 
The classification we give here is related with to Williamson's canonical forms for autonomous quadratic
Hamiltonians (\cite{Williamson1936} , also in \cite{Arnold1989}, appendix 6), but it is not 
an extension of it. Let us recall that Williamson classification is a classification under
real independent of time changes of frame. 
 
  In the list below, canonical forms are listed up to a factor of proportionality in $\overline{\mathbb M(\Gamma)}$
which does not alter the integrability class. However, we should notice that, as can be easily seen in the list, 
for certain specific values of 
$f(t)\in\overline{\mathbb M(\Gamma)}$ in the cases (3), (4) and (5), the Galois group collapses and we fall into the cases (1) or (2).

\medskip

{\noindent\bf Theorem \ref{Canonical}.}
\emph{
Let $H(t,x_1,x_2,y_1,y_2)\in \overline{\mathbb M(\Gamma)}[V]_2$ be an integrable quadratic homogeneous 
Hamiltonian of $2+\frac{1}{2}$ degrees of freedom. Then, there exist a symplectic change of
frame,
$$\left( \begin{array}{c} \xi_1 \\ \xi_2 \\ \eta_1 \\ \eta_2 \end{array}\right) =  
B(t)\left(\begin{array}{c}x_1 \\ x_2 \\ y_1 \\ y_2 \end{array}\right)$$
with $B(t)\in {\rm Sp}(4,\overline{\mathbb M(\Gamma)})$ 
such that, for the transformed Hamiltonian $\bar H(\xi_1,\xi_2,\eta_1,\eta_2)$,
$$\bar H = H - (\xi_1, \xi_2 ,\eta_1, \eta_2)J\dot BB^{-1}\left(\begin{array}{c} \xi_1 \\ \xi_2 \\ \eta_1 \\ \eta_2 \end{array}\right), 
\quad J = \left(\begin{array}{cc} & I \\ -I &  \end{array}  \right)$$
belongs to one of the following categories:
}
\begin{center}
  \begin{tabular}{| c | c | c | c |}
    \hline
    Normal Form & Galois & Quadratic Invariants & Parameters \\ \hline \hline
     $0$ & $\{1\}$ & All &  \\ \hline
    $f(t)\left(\xi_1\eta_1 + \frac{p}{q}\xi_2\eta_2\right)$ & $\mathbb C^*$ & $\xi_1\eta_1$, $\xi_2\eta_2$ & 
    $f(t),\frac{p}{q}$\\ \hline
    $f(t)(\xi_1\eta_1 + \xi_2\eta_2)$ & $\mathbb C^*$ & 
    $\xi_1\eta_1$, $\xi_2\eta_2$, $\xi_1\eta_2-\xi_2\eta_1$ & $f(t)$ \\ \hline
    $f(t)\xi_1\eta_1$ & $\mathbb C^*$ & $\xi_1\eta_1$, $\xi_2^2$, $\eta_2^2$, $\xi_2\eta_2$ & $f(t)$\\ \hline
    $f(t)\frac{\eta_1^2}{2}$ & $\mathbb C$ & $\eta_1^2$, $\xi_2^2$, $\xi_2\eta_2$, $\eta_2^2$ & $f(t)$ \\ \hline
    $f(t)\frac{\eta_1^2+\eta_2^2}{2}$  & $\mathbb C$ & $\eta_1^2$, $\eta_2^2$ & f(t)\\ \hline
    $f(t)\left(\xi_2\eta_1 + \lambda\eta_1^2+\frac{\eta_2^2}{2}\right)$ & $\mathbb C$ & 
    $2\xi_2\eta_1 + \eta_2^2$, $\eta_1^2$ & $f(t)$, $\lambda$ \\ \hline 
    $f(t)\xi_1\eta_1 + g(t)\xi_2\eta_2$ & $(\mathbb C^*)^2$ & $\xi_1\eta_1$, $\xi_2\eta_2$ & $f(t)$, $g(t)$ \\ \hline
    $f(t)\frac{\eta_1^2}{2} + g(t)\xi_2\eta_2$ & $\mathbb C\times\mathbb C^*$ & $\eta_1^2$, $\xi_2\eta_2$ &
    $f(t)$, $g(t)$ \\ \hline
    $f(t)\frac{\eta_1^2}{2} + g(t)\eta_2^2$ & $\mathbb C^2$ & $\eta_1^2$, $\eta_2^2$ & $f(t)$, $g(t)$ \\ \hline
    $f(t)\eta_1\left(\xi_2 + g(t)\eta_1+\frac{\eta_2^2}{2}\right)$ & 
    $\mathbb C^2$ &  $2\eta_1\xi_2 + \eta_2^2$, $\eta_1^2$ & 
    $f(t)$, $g(t)$ \\ \hline
  \end{tabular}
\end{center}
\emph{
Where $f(t)$ and $g(t)$ are arbitrary meromorphic functions, $\lambda$ is an arbitrary constant,
and $p,q$ are coprime integers.  }

\section{Main Concepts}

We are interested in non-autonomous Hamiltonian systems. By technical
reasons, we are going to consider the coefficients of those Hamiltonians
to be multivalued meromorphic functions of time, for which we will allow finite
ramification points. Therefore we will consider 
a Riemann surface $\Gamma$ endowed with a meromorphic derivation. In most
cases $\Gamma$ will be an open subset of the complex projective line $\overline{\mathbb C}$.
The field $\mathbb M(\Gamma)$ of meromorphic functions on $\Gamma$ is then a
differential field (see Section \ref{DifferentialGalois}) and so is its algebraic closure $\overline{\mathbb M(\Gamma)}$.
By abuse of notation, we will denote this field by $\frac{\partial}{\partial t}$ and its
dual $1$-form as $dt$, even in the case in which there is no $t\in\mathbb M(\Gamma)$ (for
instance, its happens when $\Gamma$ is a complex torus). 

The construction that we are going to give here is, in fact, suitable for for coefficients 
in any differential field of characteristic zero with algebraically closed field of constant.

Let $V$ be a symplectic vector space of complex dimension $2n$, 
endowed with a system of symplectic coordinates $x_1,\ldots,x_n$,
$y_1,\ldots, y_n$ such that the symplectic form on $V$ is written,
\begin{equation}\label{Omega2}
\Omega_2 = \sum_{i=1}^n dx_i\wedge dy_i.
\end{equation}

The ring $\mathbb M(V)$ of meromorphic functions on $V$ in then endowed with a Poisson Bracket, 
\begin{equation}\label{Poisson1}
\{F,G\} = 
 \sum_{i=1}^n(F_{x_i}G_{y_i}-F_{y_i}G_{x_i}).
\end{equation}

\subsection{Non-autonomous Hamiltonians}\label{NAHamiltonians}

We want to consider Hamiltonian systems depending on time through algebraic
functions. Therefore, we consider $\overline{\Gamma}$ to be any ramified covering of $\Gamma$
and the field $\mathbb M(\overline{\Gamma} \times V)$ of meromorphic functions on $\overline{\Gamma}\times V$. This
manifold, $\overline{\Gamma}\times V$ is not a symplectic manifold. However, we can consider
the $2$-form $\Omega_2$ of equation (\ref{Omega2}), which becomes now a degenerated
$2$-form, and the Poisson bracket induced, that we denote using a double bracket:
$$\{\{ F, G \}\} = 
\sum_{i=1}^n(F_{x_i}G_{y_i}-F_{y_i}G_{x_i}).$$

  For a given $H\in \mathbb M(\overline{\Gamma}\times V)$, there exist a unique meromorphic
vector field $\vec X_H$  in $\overline{\Gamma} \times V$ such that,  
\begin{equation}\label{defX} 
 i_{\vec X_H}\Omega_2 = H_tdt - dH, \quad \langle \vec X_H, dt \rangle = 1.
\end{equation}
We call $\vec X_H$ the \emph{Hamiltonian vector field} associated with the
\emph{Hamiltonian function} $H$. The equation for integral curves of
$\vec X_H$ are the usual Hamilton-Jacobi equations of motion:
$$\frac{dx_i}{dt} = H_{y_i}, \quad \frac{dy_i}{dt} = - H_{x_i}.$$

It is important to remark that the definition of $\vec X_H$ is sensible to
changes of frame in the bundle $\overline{\Gamma}\times V$. As it is written in formula
(\ref{defX}), the vector field $\vec X_H$ depends on the derivative of
$H$ with respect to $t$. So that, we understand that the vector field
$\frac{\partial}{\partial t}$ is given as a data of the problem. In
general, we can write the motion equation:

$$\frac{d\xi}{dt} = \{\{\xi, H\}\} + \xi_t, \quad \xi \in \mathbb M(\overline{\Gamma} \times V).$$

We are specifically interested in non-autonomous quadratic homogeneous Hamiltonians $H \in \overline{\mathbb M(\Gamma)}[V]_2$.
They are written as, 
$$H = \sum_{i=1,j=1}^n a_{ij}\frac{x_ix_j}{2} + b_{ij}\frac{y_iy_j}{2} + c_{ij}x_iy_j,$$
where $A = (a_{ij})$ and $B=(b_{ij})$ are symmetric and $C=(c_{ij})$ is square matrix
with coefficients in $\overline{\mathbb M(\Gamma)}$. 

  For such Hamiltonians, the vector field $\vec X_H$ is a meromorphic vector field
in $\overline{\Gamma}\times V$ for an suitable ramified covering $\overline{\Gamma}$ of $\Gamma$ and
the equations of motion
\begin{equation}\label{NAsystem}
\frac{d}{dt}\left( \begin{array}{c} x \\ y \end{array} \right)
=
\left(  \begin{array}{cc} C^t & B \\ -A & -C \end{array}\right) 
\left( \begin{array}{c} x \\ y \end{array} \right)
\end{equation}
form a system of $2n$ linear differential equations that can be investigated
from the standpoint of differential Galois theory. 

\subsection{Non-autonomous Integrability}

We present a definition of complete integrability for non-autonomous 
Hamiltonian systems. Let us consider $H\in \mathbb M(\overline{\Gamma}\times V)$. 
For the understanding of our definition is essential to note that
the ramified covering $\overline{\Gamma}$ of $\Gamma$ we consider is just 
geometric tool we need in order to deal with finitely-many valued functions of $t$.
Whenever new algebraic functions appear we just consider a new ramified
covering $\widehat{\Gamma}$ of $\overline{\Gamma}$ and lift up all the structure. We
recall that a ramified covering induces a natural algebraic field extension,
$\mathbb M(\overline{\Gamma}\times V) \subset \mathbb M(\widehat{\Gamma} \times V)$
compatible with all the differential calculus we are using here. 

Hence, in the following definition we should say integrability 
by functions which are \emph{meromorphic on $V$ and algebraic (finitely-many valued)
on $t$}. However, in order to make the exposition clear we will write
just \emph{integrable in the non-autonomous sense}.

\begin{defn}
  Let $H\in \mathbb M(\overline{\Gamma}\times V)$ a Hamiltonian function. We say that
$H$ is integrable in the non-autonomous sense $t$ if there exist a ramified covering
$\widehat{\Gamma}$ of $\overline{\Gamma}$ and functions $F_1,\ldots, F_n\in \mathbb M(\widehat{\Gamma}\times V)$
such that:  
\begin{enumerate}
\item[(1)] $\{\{F_i,F_j\}\} = 0$
\item[(2)] $\vec X_H F_i = 0$
\item[(3)] $F_1,\ldots, F_n$, and $t$ are functionally independent. 
\end{enumerate}
\end{defn}

We say that $F_1,\ldots, F_n$ form a complete system of first integrals
for $H$. It is well know that, by algebraic reasons, we can never have more that $n$
independent first integrals in involution. However, is some cases it is possible to find
different complete systems of first integrals for a given Hamiltonians. Non-autonomous
Hamiltonian systems with this properties are usually called \emph{superintegrable}. For an
algebraic treatment of the superintegrability of autonomous Hamiltonian systems
see \cite{Maciejewski}

\subsection{Extended autonomous System}

Let us consider $H\in \mathbb M(\overline{\Gamma}\times V)$. There is a classical
way to extend the phase space in such a way that we obtain
a new autonomous Hamiltonian system in $n+1$ degrees of freedom
which is, in many ways, equivalent to our original non-autonomous
system. When the integrability of a non-autonomous Hamiltonian system
is studied, usually it done in this way. It is the extended system
the one which is investigated. See, for instance \cite{Acosta}.

  We consider a new variable $h$ called \emph{dissipation} that can take 
any arbitrary complex value. Then, we are considering a new phase
space, $\overline{\Gamma}\times V \times \mathbb C_h$. This phase space is 
a symplectic manifold endowed with the symplectic form,
$$\widehat\Omega_2 = \Omega_2 + dt\wedge dh.$$ 
We consider the \emph{extended Hamiltonian} $\widehat{H}$ defined:
$$\widehat{H} = H + h.$$
The autonomous Hamiltonian vector field $\vec X_{\widehat H}$, 
gives us the following equation of motion,
\begin{equation}\label{HJAutonomous}
\dot x_i = \frac{\partial H}{\partial y_i},\quad \dot y_i = -\frac{\partial H}{\partial x_i},
\quad \dot h=-\frac{\partial H}{\partial t}, \quad \dot t = 1.
\end{equation}

Let us consider the natural  projection $\pi\colon \overline{\Gamma}\times V \times \mathbb C_h \to \overline{\Gamma} \times V$.
It is clear, from the equations (\ref{HJAutonomous}) that the vector field $\vec X_{\widehat{H}}$
is projectable by $\pi$ and it is projected onto $\vec X_{H}$.

On the other hand, let $\lambda$ be any complex constant value. We can consider the hypersurface
$$\mathcal H_{\lambda} = \{(x,y,t,h)\in \overline{\Gamma} \times V \times  \mathbb C_h \,\colon\, H(x,y,t) + h = \lambda \},$$
which is a constant energy hypersurface. It is also clear that $\mathcal H_{\lambda}$ is isomorphic
by $\pi$ with $V\times T$, and that this isomorphism conjugates $\vec X_{\widehat{H}}|_{\mathcal H_{\lambda}}$
with $\vec X_{H}$.

As a first test for our definition, we expect the extended autonomous system associated
to a integrable non-autonomous system to be a completely integrable Hamiltonian system. This
fact is easy to verify. 

\begin{proposition}\label{NAIntAInt}
Assume that $H$ is integrable in the non-autonomous sense, and let $\widehat{\Gamma}$
be a ramified covering of $\overline{\Gamma}$ such that there is a complete systems of first integrals
of $\vec X_H$ in $\mathbb M(\widehat{\Gamma}\times V)$. Then, $\widehat{H}$ is 
completely integrable by meromorphic functions in $\widehat{\Gamma}\times V \times \mathbb C_h$. 
\end{proposition}

\textbf{Proof.} Let us consider $F_1,\ldots, F_n$ a complete system of first integrals in involution of
$\vec X_H$. Then, it is clear that $\hat H, F_1, \ldots, F_n$ are in involution, and they
are meromorphic functionally independent functions in $\widehat{\Gamma} \times V \times  \mathbb C_h$. \hfill $\square$

\medskip

However, the reciprocal does not hold in general. Let us assume that $\vec X_{\widehat{H}}$ is 
completely integrable by meromorphic functions. Then, there exist a complete system of first integrals including $\widehat{H}$,
take $F_1,\ldots, F_n, \widehat{H}$. We take the energy level hypersurface $\mathcal H_\lambda$.
We can construct first integrals of $\vec X_{H}$ easily by restricting the fist integrals we had to
$\mathcal H_{\lambda}$ which is isomorphic to $\bar\Gamma \times V$. We get the first integrals,  
$$\overline{F}_i(x,y,t) = F_i(x,y,t,\lambda - H(x,y,t)).$$
Then we have,
$$\{\{\overline{F}_i,\overline{F}_j\}\} = \{\{F_i , F_j \}\} + \frac{\partial F_j}{\partial h}\{\{F_i,H\}\} + 
\frac{\partial F_i}{\partial h}\{\{F_j, H\}\},$$
and therefore the $\overline{F}_i$ are not in general in involution. In spite of it, we can can state 
the reciprocal for the case of Hamiltonians of $1+\frac{1}{2}$ degrees of freedom, in which the 
hypothesis of involution is not necessary.

\begin{theorem}
  Let be $H\in \mathbb M(\overline{\Gamma} \times \mathbb C^2_{x,y})$ a non-autonomous Hamiltonian of $1+\frac{1}{2}$ 
degrees of freedom. Then $H$ is integrable in the non-autonomous sense by meromorphic functions in
$\widehat{\Gamma}\times \mathbb C^2_{x,y}$ if and only if its associated autonomous extended Hamiltonian $\widehat H = H+h$ 
is completely integrable by meromorphic functions in $\widehat{\Gamma}\times \mathbb C^2_{x,y}\times \mathbb C_h$. 
\end{theorem}

\section{Differential Galois Theory}\label{DifferentialGalois}

The differential Galois theory deals with the integrability
by quadratures of systems of linear differential equations. 
In this section we will develop only the part of the theory
we need for our purposes, and we will give no proofs of the
facts we expose. The interested reader may consult
more complete references like \cite{Kaplansky, Kolchin, MoralesLibro, Blazquez}.

Let $\mathbb K$ be a field of characteristic zero. A derivation in $\mathbb K$ is an
additive map $\partial \colon \mathbb K \to \mathbb K$ which satisfies the Leibniz rule
$$\partial(ab) = b\partial (a) + a\partial(b).$$

A \emph{differential field} is a pair $(\mathbb K,\partial_{\mathbb K})$ consisting on a field
and a derivation on it. By abuse of notation we will write ${\mathbb K}$ instead of
the pair $(\mathbb K,\partial_{\mathbb K})$ whenever it does not lead to confusion. 

Given a differential field $\mathbb K$, we denote by $C(\mathbb K)$ the field
of \emph{constants} of $\mathbb K$ which consists of the elements $a\in\mathbb K$ 
such that $\partial_{\mathbb K} a$ vanish. From now on we will consider always a differential field
$\mathbb K$ whose field of constants is an algebraically closed field $\mathbb C$ of characteristic
zero. In most examples we are going to consider, this field of constants is the field of the complex numbers.

\begin{example}
Let us consider a non-autonomous Hamiltonian $H \in \mathbb M(V\times T)$,
as in subsection \ref{NAHamiltonians}. Then, the field $\mathbb M(V\times T)$
endowed with the derivation,
$$\partial F = - \{\{ H, F\}\} = \vec X_{H}F $$
is a differential field. Its field of constants consist of the meromorphic first
integrals of the Hamiltonian vector field $\vec X_H$. 
\end{example}

\subsection{Picard-Vessiot Extensions}

An extension of differential fields $\mathbb K\subset \mathbb L$ is an inclusion of $\mathbb K$ in $\mathbb L$
which is an extension of fields and $\partial_{\mathbb L}|_{\mathbb K} = \partial_{\mathbb K}$. In what follows
all extensions of fields we consider are differential extensions.

\begin{example}
Let $\overline{\mathbb K}$ be the algebraic closure of $\mathbb K$. Then, the derivation extends to 
$\overline{\mathbb K}$ in a unique way and $\mathbb K \subset \overline{\mathbb K}$ is a differential
field extension. Furthermore, $C(\overline{\mathbb K})= \mathbb C$ since $\mathbb C$ is algebraically
closed.
\end{example}

Let $\mathbb K \subset \mathbb L$ be a differential field extension. A differential automorphism
of $\mathbb L$ over $\mathbb K$ is a field automorphism $\sigma$ of $\mathbb L$ which commutes
with the derivation and fix $\mathbb K$ point-wise. The set of all differential automorphisms
of $\mathbb L$ over $\mathbb K$ is clearly a group that we denote by ${\rm Aut}_{\mathbb K}(\mathbb L)$.

Let us consider a system of linear homogeneous differential equations with coefficients in $\mathbb K$,
\begin{equation}\label{LinearHomogeneous}
y' = Ay, \quad A\in \mathfrak{gl}(n,\mathbb K)
\end{equation}
and an extension $\mathbb K \subset \mathbb E$.  The set
of solution of (\ref{LinearHomogeneous}) in $\mathbb E^n$ form 
a subset $S_{\mathbb E}\subset \mathbb E^n$ which is a vector space over
$C(\mathbb E)$ of dimension lee or equal than $n$. We denote
by $\mathbb K(S_{\mathbb E})$ to the smallest differential field containing both $\mathbb K$ and
the coordinates of elements of $S_{\mathbb E}$.
  
A differential field extension $\mathbb K\subset\mathbb L$ is called a \emph{Picard-Vessiot}
extension for (\ref{LinearHomogeneous}) if the following conditions hold:
\begin{itemize}
\item[(1)] There is a fundamental matrix of solutions of (\ref{LinearHomogeneous}) in ${\rm GL}(n,\mathbb L)$.
\item[(2)] $\mathbb L$ is generated over $\mathbb K$ by the solutions of (\ref{LinearHomogeneous}), id est, $\mathbb L = \mathbb K(S_{\mathbb L})$.
\item[(3)] There is no new constants in $\mathbb L$, $C(\mathbb L)=\mathbb C$. 
\end{itemize}

Any system of linear homogeneous differential equations as (\ref{LinearHomogeneous}) admits
a Picard-Vessiot extension. Those extensions are unique up to an isomorphism that fixes
$\mathbb K$ point-wise. Therefore, we will speak on \emph{the Picard-Vessiot extension
associated with} (\ref{LinearHomogeneous}) .

In general, a differential field extension  $\mathbb K \subset \mathbb L$ is called a Picard-Vessiot 
extension if it is a Picard-Vessiot for certain system of linear homogeneous equations
with coefficients in $\mathbb K$.
If $\mathbb K \subset \mathbb L$ is a Picard-Vessiot extension then for any intermediate
extension $\mathbb K \subset \mathbb L_1 \subset \mathbb L$ we have that $\mathbb L_1 \subset \mathbb L$
is also a Picard-Vessiot extension. A remarkable property of Picard-Vessiot extensions
is the normality, for any $a\in\mathbb L$ not in $\mathbb K$ there exist an automorphism 
$\sigma\in{\rm Aut}_{\mathbb K}(\mathbb L)$ such that $\sigma(a)\neq a$. 

\subsection{Differential Galois Group}

Let $\mathbb K\subset \mathbb L$ a Picard-Vessiot extension with common field of constants $\mathbb C$
for the system (\ref{LinearHomogeneous}). Let us consider 
$S_{\mathbb L}\subset \mathbb L^n$ the set of solutions of such system. It is a $\mathbb C$-vector space. 
Any differential automorphism $\sigma$ fix point-wise $\mathbb K$ and then let the equations (\ref{LinearHomogeneous})
invariant. Therefore, it induces a $\mathbb C$-linear map $\phi_{\sigma}\colon V\to V$.
This gives us a faithful representation,
$${\rm Aut}_{\mathbb K}(\mathbb L) \to {\rm GL}(S_{\mathbb L},\mathbb C).$$

The image of this map is called the \emph{differential Galois group} of the system  (\ref{LinearHomogeneous}). By
abuse of notation it will be denoted by ${\rm Gal}_{\mathbb L/\mathbb K}$. However, let us recall
that this group, is associated to the equation, not to the differential field extension.

 The most remarkable property of ${\rm Gal}_{\mathbb L/\mathbb K}$
is that it is an \emph{linear algebraic subgroup} of ${\rm GL}(S_{\mathbb L},\mathbb C)$. A linear
algebraic subgroup is just a matrix group defined by polynomial equations in the matrix
elements. In linear algebraic groups is natural to consider the Zariski topology, for what
closed subsets are defined by polynomial equations. With this topology any algebraic group has
a finite number of connected components and the connected component which contains the identity
is the biggest normal algebraic subgroup of finite index.

  Let $G\subset {\rm Gal}_{\mathbb L/\mathbb K}$ be an algebraic subgroup. We can assign to it
the intermediate extension $\mathbb K \subset \mathbb L^G \subset \mathbb L$, being $\mathbb L^G$
the field of elements that are fixed by $G$. Reciprocally, for any intermediate extension
$\mathbb K \subset \mathbb F \subset \mathbb L$ the group $G_{\mathbb F}$ of automorphisms of
$\mathbb L$ that fix $\mathbb F$ point-wise is an algebraic subgroup $G\subset {\rm Gal}_{\mathbb L/\mathbb K}$. 

\begin{proposition}\label{GaloisCorrespondence}
  The assignation $G\leadsto \mathbb L^G$ gives a one to one correspondence between the set of 
all algebraic subgroups of  ${\rm Gal}_{\mathbb L/\mathbb K}$ and intermediate extensions
$\mathbb K \subset \mathbb L$. A subgroup $G$ is normal in ${\rm Gal}_{\mathbb L/\mathbb K}$ 
if and only if $\mathbb K \subset \mathbb L^G$ is a Picard-Vessiot extension. In such case its
Galois group is ${\rm Gal}_{\mathbb L/\mathbb K}/G$ 
\end{proposition}

Note that a Picard-Vessiot extension is algebraic if only if its Galois group
is finite and purely transcendental if and only if its Galois group is connected.
Let us denote ${\rm Gal}_{\mathbb L/\mathbb K}^0$ to the connected component of 
the identity of the Galois group of $\mathbb L$ over $\mathbb K$. Using the
Proposition \ref{GaloisCorrespondence} we can split out the extension in
two Picard-Vessiot extensions,
$$\mathbb K \subset \mathbb K^0 \subset \mathbb L$$
such that $\mathbb K \subset \mathbb K^0$ is an algebraic extension with
finite Galois group ${\rm Gal}_{\mathbb L/\mathbb K}/{\rm Gal}_{\mathbb L/\mathbb K}^0$
and $\mathbb K^0 \subset \mathbb L$ is purely transcendental with connected Galois
group ${\rm Gal}_{\mathbb L/\mathbb K}^0$. 

In particular, if $\mathbb K$ is algebraically closed, any Picard-Vessiot
extension has connected Galois group.

For an algebraic subgroup $G\subset {\rm GL}(n,\mathbb C)$ with Lie algebra
$\mathfrak{g} \subset \mathfrak{gl}(n,\mathbb C)$ and any field extension
$\mathbb C \subset\mathbb F$ we will denote by $G(\mathbb F)$ the algebraic subgroup of
${\rm GL}(n,\mathbb F)$ defined by the same equations and by $\mathfrak{g}(\mathbb F)$
its Lie algebra. The Galois group of a system (\ref{LinearHomogeneous}) is bounded by the matrix of coefficients. 
If this matrix takes values in a fixed Lie subalgebra of $\mathfrak{gl}(n,\mathbb C)$ the Galois group can not
growth beyond. The following result is well known and can be found in \cite{Blazquez}.

\begin{proposition}
Let $G\subset {\rm GL}(n, \mathbb C)$ be an algebraic subgroup of ${\rm GL}(n,\mathbb C)$ and let $\mathfrak g$
its Lie algebra. Let us consider a system of equations,
\begin{equation}\label{LinearHomogeneous2}
y' = Ay, \quad A\in \mathfrak{g}(\mathbb K)
\end{equation}
such that the matrix of coefficients $A$ relies in the Lie algebra of $G$, and
$\mathbb K\subset \mathbb L$ its Picard-Vessiot extension. 
We can take a basis of $S_{\mathbb L}$, the space of solutions of {\rm (\ref{LinearHomogeneous2})} in $\mathbb L^n$,
in such way that,
$${\rm Gal}_{\mathbb L/\mathbb K} \subset G$$
\end{proposition}

\subsection{Lie-Kolchin Reduction}

Let us consider the system (\ref{LinearHomogeneous}), and $\mathbb K\subset\mathbb K_1$
a field extension with no new constants. We can take take a change of variables,
$z = By$, with $B\in {\rm GL}(n, \mathbb K_1)$ obtaining a new equation for $z$,
\begin{equation}\label{COF}
z' = (B'B^{-1} + BAB^{-1})z
\end{equation}
where the new matrix of coefficients $(B'B^{-1} + BAB^{-1})$ is now in $\mathfrak{gl}(n,\mathbb K_1)$.

The connected component of the identity of the Galois group represent the smallest subgroup to 
which our differential equation can be reduced by means of a change of variables involving
algebraic functions. The following result if due to Kolchin and Kovacic \cite{Kolchin}, and is very
close to a method of reduction of differential equations due to Lie. For a modern presentation,
and comparison between those two see \cite{Blazquez}.

\begin{theorem}\label{KolchinReduction}
Let us consider the system {\rm (\ref{LinearHomogeneous2})} and its associated Picard-Vessiot extension 
$\mathbb K\subset\mathbb L$. Let us fix any basis of $S_{\mathbb L}$, the space of solutions of (\ref{LinearHomogeneous})
in $\mathbb L^n$ so that we get a representation of ${\rm Gal}_{\mathbb L/\mathbb K}$ in $G$.
Let $\overline{\mathbb K}$ be the algebraic closure of $\mathbb K$. Then, there exist $B\in G(\overline{\mathbb K})$
such that $\bar A = B'B^{-1} + BAB^{-1}$ is in the Lie algebra of the Galois group $\mathfrak{gal}_{\mathbb L/\mathbb K}(\overline{\mathbb K})\subset 
\mathfrak{g}(\overline{\mathbb K})$:
$$z = By, \quad z' = \bar Az, \quad \bar A \in \mathfrak{gal}_{\mathbb L/\mathbb K}(\overline{\mathbb K}).$$
\end{theorem}

\subsection{Morales-Ramis Integrability Condition}

Let $V$ be a complex symplectic manifold of dimension $2n$, with symplectic form $\Omega_2$ that we write in local coordinates,
$$\Omega_2 = \sum_{i=1}^n dx_i\wedge dy_i.$$
The field of meromorphic functions $\mathbb M(V)$ is then endowed with a Poisson bracket as in (\ref{Poisson1})
and to any function $H\in \mathbb M(V)$ its correspond a Hamiltonian vector field $\vec X_H$. 

Let us recall that two functions $F$, $G$ are said to be \emph{in involution} if $\{F, G\}$ vanish, and a
Hamiltonian  $H\in \mathbb M(V)$ is called completely integrable by meromorphic functions in $V$ if there
exist $n$ first integrals of $\vec X_H$, $F_1,\ldots,F_n$, functionally independent and in involution. 

Let $\Gamma$ be an integral curve of $\vec X_H$. Let us consider $\mathbb K$ to be the field of meromorphic 
functions on $\Gamma$. The vector field $\vec X_H$ gives to $\mathbb K$ the structure of a differential field.
We consider the first variational equation of $\vec X_H$ along $\Gamma$ as a system of linear differential equations
with coefficients in $\mathbb K$. This equation can be written as follows:
\begin{equation}\label{EV}
\left( \begin{array}{c} \xi_i' \\ \eta_i' \end{array} \right)  = 
\left( \begin{array}{cc} \frac{\partial H}{\partial y_ix_j}|_\Gamma & \frac{\partial H}{\partial y_iy_j}|_\Gamma \\ 
-\frac{\partial H}{\partial x_ix_j}|_\Gamma & -\frac{\partial H}{\partial x_iy_j}|_\Gamma \end{array}\right) 
\left( \begin{array}{c} \xi_i \\ \eta_i \end{array} \right).
\end{equation}

This equation has always a known solution given by the Hamiltonian vector field, $\left(\frac{\partial H}{\partial y_i}|_\Gamma, 
-\frac{\partial H}{\partial y_i}|_\Gamma\right)$ and a know invariant $\frac{\partial H}{\partial x_i}\xi_i + \frac{\partial H}{\partial y_i}\eta_i$
given by $dH|_\Gamma$. They allow us to reduce the system (\ref{EV}) to a system of dimension $2(n-1)$ called
the \emph{normal variational equation}.

Let us recall that an algebraic group is called \emph{virtually abelian} if an only if its Lie algebra is abelian, if and only if its
connected component of the identity is abelian. The following result \cite{MoralesRamis1, MoralesLibro} gives us 
an algebraic criterium for the complete integrability of Hamiltonian systems. 

\begin{theorem}\label{MoralesRamisTheorem}
Assume that $H$ is completely integrable by terms of meromorphic first integrals. Let $\Gamma$ be an integral curve of
$\vec X_H$. Then the Galois groups of the variational equation and the normal variational equation of $\vec X_H$ along $\Gamma$ 
are virtually abelian. 
\end{theorem}

\section{Linear non-autonomous Hamiltonian Systems}

Let us consider $V$ a $2n$-dimensional symplectic vector space over $\mathbb C$ with
symplectic form $\Omega_2 = \sum_{i=1}^ndx_i\wedge dy_i$ and $\Gamma$ a Riemann surface 
endowed with a meromorphic derivation.  We have seen that for any 
quadratic homogeneous non-autonomous Hamiltonian $H\in \overline{\mathbb M(\Gamma)}[V]_2$
the equations of motion take the form given by formula (\ref{NAsystem}).

Let us consider the matrix, 
$$J = \left(\begin{array}{cc} 0 & I \\ -I & 0 \end{array} \right),$$
where $I$ denotes the identity matrix of rank $n$. We define the
symplectic group ${\rm Sp}(2n,\mathbb C)$ to the the group of all non-degenerate matrices $\sigma$
such that $\sigma^tJ\sigma = J$. It is an algebraic subgroup of ${\rm GL}(2n,\mathbb C)$.
Its lie algebra $\mathfrak{sp}(2n,\mathbb C)$ consist in all matrices $A$ such that
$A^tJ + JA = 0$. It is clear that the matrix of coefficients of (\ref{NAsystem})
is in $\mathfrak{sp}(2n,\overline{\mathbb C(t)})$, and then the Galois group of
such equation is a subgroup of ${\rm Sp}(2n,\mathbb C)$.

\subsection{Lie Algebra Structure}

Let us consider $\mathbb C[V]_2$ the space of quadratic homogeneous polynomials
on the linear coordinates of $V$ with complex coefficients. Then, the assignation:
\begin{equation}\label{LiePoisson}
\mathbb C[V]_2 \xrightarrow{\quad \sim\quad} \mathfrak{sp}(2n,\overline{\mathbb M(\Gamma)})
\end{equation}
$$\sum_{i=1}^n \left(\frac{a_{ij}}{2}x_ix_j + \frac{b_{ij}}{2}y_iy_j + c_{ij}x_iy_j\right) \to \left(\begin{array}{cc} C^t & B \\ -A & -C \end{array} \right) 
$$
brings us an isomorphism between the Poisson structure of $\mathbb C[V]_2$ and the
Lie algebra structure of $\mathfrak{sp}(2n,\overline{\mathbb M(\Gamma)})$.
This fact is well known (see for instance \cite{MoralesLibro}, page 64).

\subsection{Changes of Frame}

Quadratic homogeneous Hamiltonian behave nicely with respect to linear
changes of frame. Let us take a new system of coordinates,
$$\left( \begin{array}{c} \xi \\ \eta  \end{array}\right) =  
B(t)\left(\begin{array}{c}x \\ y \end{array}\right)$$
with $B(t)\in {\rm Sp}(4,\overline{\mathbb M(\Gamma)}).$ 
Then, by using (\ref{COF}) and (\ref{LiePoisson}) we can write down a \emph{transformed Hamiltonian},
$$\bar H = H - (\xi ,\, \eta)J\dot BB^{-1}\left(\begin{array}{c} \xi \\ \eta \end{array}\right), 
\quad J = \left(\begin{array}{cc} & I \\ -I &  \end{array}  \right),$$
which gives the equations of the movement in the new system of coordinates,
$$\dot\xi = \{\xi, \bar H \}, \quad \dot \eta = \{\eta, \bar H\}.$$ 

\subsection{Integrability}

\begin{lemma}\label{NVEitself}
Let $H\in \overline{\mathbb M(\Gamma)}[V]_2$ be a quadratic homogeneous non-autonomous Hamiltonian. Let
us consider $\vec X_{\widehat H}$ the associated extended autonomous vector field. 
Then, the normal variational equation of $\vec X_{\widehat H}$ along 
any integral curve $\Gamma$ coincides with $\vec X_H$ itself. 
\end{lemma}

{\bf Proof.} 
Let $\widehat{H} = \frac{a_{ij}}{2} x_ix_j + \frac{b_ij}{2}y_iy_j + c_{ij}x_iy_j + h$. Then, the variational equation
of $\vec X_{\widehat{H}}$ around an integral curve $\Gamma$ is:
$$\left(\begin{array}{c} \xi'\\  \eta' \\ \tau' \\ \chi'  \end{array}\right)=
\left(\begin{array}{cccccccc} 
C^t & B & H_{{y_i}t}|_\Gamma &  0 \\ 
-A & - C &  -H_{{x_i}t}|_\Gamma & 0  \\
0  & 0 & 0 & 0 \\
- H_{t{x_j}}|_\Gamma & -H_{t{y_j}}|_{\Gamma} & -H_{tt}|_\Gamma &  0 
\end{array}\right)
\left(\begin{array}{c} \xi\\  \tau \\ \eta \\ \chi  \end{array}\right)
$$
The Normal Variational Equation is obtained by using the know solution $\tau = 1,$ $\xi = \chi = \eta = 0$
and restriction to the hyperplane $\chi=0$. We easily get,
$$\left(\begin{array}{c} \xi'  \\ \eta'  \end{array}\right)=
\left(\begin{array}{cccccccc} 
C^t &  B \\ 
-A & -C  \\
\end{array}\right)
\left(\begin{array}{c} \xi\\   \eta  \end{array}\right),
$$
which gives us the normal variational equation of the statement. 


\begin{theorem}\label{main}
Let $H\in \overline{\mathbb M(\Gamma)}[V]_2$ be a quadratic homogeneous non-autonomous Hamiltonian
of $2+\frac{1}{2}$ degrees of freedom, with coefficients meromorphic in $\overline{\Gamma}$ a ramified
covering of $\Gamma$. The following are equivalent:
\begin{enumerate}
\item[(1)] The associated extended autonomous system $\widehat{H}$ is completely
integrable by meromorphic functions in $\widehat{\Gamma} \times V\times \mathbb C_h$ for some
ramified covering $\widehat{\Gamma}$ of $\overline{\Gamma}$.
\item[(2)] $H$ is integrable in the non-autonomous sense  by meromorphic functions 
in $\\widehat{\Gamma} \times V$ for some ramified covering $\widehat{\Gamma}$ of $\overline{\Gamma}$.
\item[(3)] $H$ is integrable in the non-autonomous sense by quadratic first
integrals $F_1,F_2\in \overline{\mathbb M(\Gamma)}[V]_2$.
\item[(4)] The connected component of the Galois group of $\vec X_{H}$ is a abelian.
\end{enumerate}
\end{theorem}

{\bf Proof}:

\smallskip
 $(4)\Rightarrow (3)$ 
 This part of the proof relies in the classification of connected abelian subgroups of the symplectic
which is made in section \ref{Class}. 
 By Lie-Kolchin reduction, Theorem \ref{KolchinReduction}, there
exist a symplectic change of frame $B\in Sp(4,\overline{\mathbb M(\Gamma)})$ such that,
\begin{equation}\label{CHF}
\left(\begin{array}{c} \xi_1 \\ \xi_2 \\ \eta_1 \\ \eta_2 \end{array} \right) = B\left(\begin{array}{c} x_1 \\ x_2 \\ y_1 \\ y_2 \end{array}\right) 
\end{equation}
and
\begin{equation}\label{THS}
\left(\begin{array}{c} \xi_1' \\ \xi_2' \\ \eta_1' \\ \eta_2' \end{array} \right) = A\left(\begin{array}{c} \xi_1 \\ \xi_2 \\ \eta_1 \\ \eta_2 \end{array} \right),
\quad A\in \mathfrak{g}(\overline{\mathbb M(\Gamma)}).
\end{equation}
Where $\mathfrak g$ is an abelian subalgebra of $\mathfrak{sp}(2,\mathbb C)$. By Corollary \ref{Dim2}, $\mathfrak g$ 
is contained in an abelian subalgebra, spanned by two linear Hamiltonian vector fields $\vec Y_1$ and $\vec Y_2$
with constants coefficients. By the dictionary between $\mathfrak{sp}(2,\mathbb C)$ and $\mathbb C[V]_2$ there are
two quadratic polynomials $F_1, F_2$ such that $\vec Y_1 = \vec X_{F_1}$ and $\vec Y_2 = \vec X_{F_2}$. Then,
its is clear that  $F_1(\xi_1,\xi_2,\eta_1,\eta_2)$ and $F_2(\xi_1,\xi_2,\eta_1,\eta_2)$ are first integrals of 
(\ref{THS}) in involutions. Substituting these variables using (\ref{CHF}) we get two first integrals of 
$\vec X_H$ in involution in $\overline{\mathbb M(\Gamma)}[V]_2$. 

\smallskip
$(3)\Rightarrow (2)$ In particular the first integrals in $\overline{\mathbb M(\Gamma)}[V]_2$ are meromorphic
functions in $\widehat{\Gamma}\times V$ for some covering $\widehat{\Gamma}$ of $\Gamma$.

\smallskip
$(2)\Rightarrow (1)$ It is Proposition \ref{NAIntAInt}.

\smallskip
$(1) \Rightarrow (4)$ By Lemma \ref{NVEitself}, we have that $\vec X_H$ coincides with the normal
variational equation of its associated extended autonomous system $\vec X_{\widehat H}$ along any particular solution.
We take $\Gamma$ the particular solution corresponding to $x_i=0$, $y_i=0$, $h=0$, $t=t$,
therefore the field of coefficients is here $\mathbb M(\overline{\Gamma})$ which is an algebraic extension of $\mathbb M(\Gamma)$. 
By Theorem \ref{MoralesRamisTheorem}, we have that the Galois group of this equation is virtually abelian. It finish the proof. 
\hfill $\square$

\begin{remark}
The proof of the above systems relies on Corollary \ref{Dim2} which is proven in the next section. It is well known that
any abelian subalgebra of $\mathfrak{sp}(2n, \mathbb C)$ has dimension at most $n$, and there are maximal abelian subalgebras
of dimension $n$. However, it is beyond the knowledge of the authors if any maximal abelian subalgebra realizes dimension $n$.
If this holds, then Theorem \ref{main} will hold for any number of degrees of freedom.  
\end{remark}

\section{Classification of Connected Abelian Subgroups of ${\rm Sp}(4,\mathbb C)$}\label{Class}

In this section we classify the connected abelian subgroups of  ${\rm Sp}(4,\mathbb C)$. In the paper
\cite{Churchill1995} there are shown the canonical form of \emph{$2$-Ziglin} subgroups that
lead to certain obstructions to integrability. In a similar way, the classification of connected abelian connected 
subgroups completes the proof of Theorem \ref{main} and allow us to compute canonical forms for integrable system.

Let us recall that any connected abelian subgroups of ${\rm Sp}(4,\mathbb C)$ must be of dimension one or two, and that
any connected abelian linear group is direct product of multiplicative $\mathbb C^*$ and addictive $\mathbb C$ groups.  
The following technical lemma can be proved by direct computation. 

All computations of this section are easy to reproduce, so that we will give just sketchs of the proofs. 

\begin{lemma}\label{nihilpotent}
Let $A$ be a nihilpotent matrix in ${\mathfrak sp}(4,\mathbb C)$ then one of the following 
conditions holds:
\begin{itemize}
\item[(1)] $A^2 = 0$, $\ker(A)$ is of dimension 3, and $A$ is conjugated to a
matrix of the form:
$$\left( \begin{array}{cccc} 
0 & 0 & 1 & 0  \\
0 & 0 & 0 & 0 \\
0 & 0 & 0 & 0 \\
0 & 0 & 0 & 0 \end{array} \right)$$

\item[(2)] $A^2 = 0$, $\ker(A)$ is of dimension 2, and $A$ is conjugated to a
matrix of the form:

$$\left( \begin{array}{cccc} 
0 & 0 & 1 & 0  \\
0 & 0 & 0 & 1 \\
0 & 0 & 0 & 0 \\
0 & 0 & 0 & 0 \end{array} \right)$$

\item[(3)]  $A^3 \neq 0$, $\ker(A)$ is of dimension 1, and $A$ is conjugated to a
matrix of the form:
$$\left( \begin{array}{cccc} 
0 & 1 & \lambda & 0  \\
0 & 0 & 0 & 1 \\
0 & 0 & 0 & 0 \\
0 & 0 & -1 & 0 \end{array} \right)$$
with $\lambda\in\mathbb C$.
\end{itemize}

\end{lemma}

{\bf Proof.} Let $A$ be a nihilpotent matrix in $\mathfrak{sp}(4,\mathbb C)$. It is well know that there exist a non-degenerate matrix $B$
such that,
$$\bar A = B^{-1}AB = \left( \begin{array}{cccc} 
0 & a & 0 & 0  \\
0 & 0 & b & 0 \\
0 & 0 & 0 & c \\
0 & 0 & 0 & 0 \end{array} \right),$$
with $a$, $b$ and $c$ equal to $0$ or $1$. In this base, the matrix of the symplectic metric is a 
non-degenerate skew-symmetric matrix,
$$\bar J = B^tJB = \left( \begin{array}{cccc} 
0 & j_{12} & j_{13} & j_{14}  \\
-j_{12} & 0 & j_{23} & j_{24} \\
-j_{13} & -j_{23} & 0 & j_{34} \\
-j_{14} & -j_{24} & -j_{34} & 0 \end{array} \right)$$
Applying that $\bar A^t\bar J + \bar J \bar A=0$ we see that the cases $a=b=1$, $c=0$ and
$a=0$, $b=c=1$ lead $|\bar J|=0$. Then, it is clear that these cases
can not happen for a symplectic matrix $A$. The cases $a=1$,$b=c=0$; $a=c=0$, $b=1$ and
$a=b=0$, $c=1$ are clearly equivalent and lead to the case (1) of the statement.
The cases $a=c=1$, $b=0$ lead to the case (2) of the statement. And the case
$a=b=c=1$ leads to the case (3) of the statement, just by applying $\bar A^t\bar J + \bar J \bar A=0$ 
and looking for the matrices that conjugate $\bar J$ to canonical form $J$.\hfill$\square$

\begin{proposition}\label{Group1}
Let $G$ be a $1$-dimensional connected abelian subgroup of ${\rm Sp}(4,\mathbb C)$,
then $G$ is conjugated to one of the following list:
\begin{itemize}
\item[(1)] Case $G$ isomorphic to the multiplicative group $\mathbb C^*$.
\begin{itemize}
\item[(1.a)]
$$ G \equiv \left\{\left( \begin{array}{cccc} \lambda^p & 0 & 0 & 0  \\
0 & \lambda^q & 0 & 0 \\
0 & 0 & \lambda^{-p} & 0 \\
0 & 0 & 0 & \lambda^{-q} \end{array} \right) \quad\colon\quad \lambda\in\mathbb C^*\right\}
$$
with $(p,q)$ relative primes. 
\item[(1.b)]
$$ G \equiv \left\{\left( \begin{array}{cccc} \lambda & 0 & 0 & 0  \\
0 & \lambda & 0 & 0 \\
0 & 0 & \lambda^{-1} & 0 \\
0 & 0 & 0 & \lambda^{-1} \end{array} \right) \quad\colon\quad \lambda\in\mathbb C^*\right\}
$$
\item[(1.c)]
$$ G \equiv \left\{\left( \begin{array}{cccc} \lambda & 0 & 0 & 0  \\
0 & 1 & 0 & 0 \\
0 & 0 & \lambda^{-1} & 0 \\
0 & 0 & 0 & 1 \end{array} \right) \quad\colon\quad \lambda\in\mathbb C^*\right\}
$$
\end{itemize}
\item[(2)] Case $G$ isomorphic to the additive group $\mathbb C$. 
\begin{itemize}
\item[(2.a)]
$$ G \equiv \left\{\left( \begin{array}{cccc} 
1 & 0 & \lambda & 0  \\
0 & 1 & 0 & 0 \\
0 & 0 & 1 & 0 \\
0 & 0 & 0 & 1 \end{array} \right) \quad\colon\quad \lambda\in\mathbb C\right\}
$$
\item[(2.b)]
$$G \equiv \left\{\left( \begin{array}{cccc} 
1 & 0 & \lambda & 0  \\
0 & 1 & 0 & \lambda \\
0 & 0 & 1 & 0 \\
0 & 0 & 0 & 1 \end{array} \right) \quad\colon\quad \lambda\in\mathbb C\right\}
$$
\item[(2.c)]
$$G \equiv \left\{ \left( \begin{array}{cccc} 1 & \lambda & -\frac{\lambda^3}{6} + k\lambda & \frac{\lambda^2}{2}  \\
0 & 1 & -\frac{\lambda^2}{2} & \lambda \\
0 & 0 & 1 & 0 \\
0 & 0 & -\lambda & 1 \end{array} \right) \quad\colon\quad \lambda\in\mathbb C\right\}
$$
with $k\in\mathbb C$. 
\end{itemize}
\end{itemize}
\end{proposition}

{\bf Proof.} 
Multiplicative groups are always contained into a maximal torus and then
they are classified. See for instance \cite{Humphreys}. Then, we have just
to classify additive group. A one dimensional subgroup of ${\rm GL}(n,\mathbb C)$
is isomorphic to the  additive group if and only if its Lie algebra is generated
by a nihilpotent matrix. Conjugacy classes of nihilpotent matrix are given
by Lemma \ref{nihilpotent}. Cases 2. a, b, and c are given just by the exponential
of these nihilpotent matrices. \hfill $\square$

\begin{theorem}\label{Group2}
Let $G$ be a maximal connected abelian subgroup of ${\rm Sp}(4,\mathbb C)$,
then $G$ is conjugated to one of the following list:
\begin{itemize}
\item[(3)] Case $G$ isomorphic to $\mathbb C^*\times \mathbb C^*$.
$$ G \equiv \left\{\left( \begin{array}{cccc} \lambda & 0 & 0 & 0  \\
0 & \mu & 0 & 0 \\
0 & 0 & \lambda^{-1} & 0 \\
0 & 0 & 0 & \mu^{-1} \end{array} \right) \quad\colon\quad \lambda,\mu\in\mathbb C^*\right\}
$$
\item[(4)] Case $G$ isomorphic to $\mathbb C\times \mathbb C^*$.
$$ G \equiv \left\{
\left( \begin{array}{cccc} 1 & 0 & \lambda & 0  \\
0 & \mu & 0 & 0 \\
0 & 0 & 1 & 0 \\
0 & 0 & 0 & \mu^{-1} \end{array} \right)
 \quad\colon\quad \lambda\in \mathbb C,\,\mu\in\mathbb C^*\right\}
$$

\item[(5)] Case $G$ isomorphic to $\mathbb C\times \mathbb C$.
\begin{itemize}
\item[(5.a)]
$$ G \equiv \left\{
\left( \begin{array}{cccc} 1 & 0 & \lambda & 0  \\
0 & 1 & 0 & \mu \\
0 & 0 & 1 & 0 \\
0 & 0 & 0 & 1 \end{array} \right)
 \quad\colon\quad \lambda,\,\mu\in\mathbb C\right\}
$$

\item[(5.b)]
$$ G \equiv \left\{
\left( \begin{array}{cccc} 1 & \lambda & \mu -\frac{\lambda^3}{6} & \frac{\lambda^2}{2}  \\
0 & 1 & -\frac{\lambda^2}{2} & \lambda \\
0 & 0 & 1 & 0 \\
0 & 0 & -\lambda & 1 \end{array} \right)
 \quad\colon\quad \lambda,\,\mu\in\mathbb C\right\}
$$

\end{itemize}
\end{itemize}
\end{theorem}

{\bf Proof.} First, let us see that any one dimensional subgroup of ${\rm Sp}(4,\mathbb C)$ is contained in a two dimensional 
abelian subgroup. If our group is multiplicative, this result is well know. If our group is additive just note that cases
2.a and 2.b are included into case 5.a here and case 2.c is included into case 5.b. 

Second, let us see that any $2$-dimensional abelian subgroup  of ${\rm Sp}(4,\mathbb C)$ falls in one of the cases we list
above. Let $G$ such a group. Then, it is isomorphic to $\mathbb C^*\times \mathbb C^*$, $\mathbb C\times \mathbb C^*$ or
to $\mathbb C\times \mathbb C$. In the first case, it is a maximal torus and it is well known that it falls
into case 3. 

In the second case, let us consider $G$ isomorphic  to $\mathbb C\times \mathbb C^*$. Let
$\mathfrak g$ be its Lie algebra. It is clear that there is a unique line in $\mathfrak g$ spanned by a nihilpotent
matrix, since there is only one algebraic morphism from $\mathbb C$ into $G$. Let $A$ be such a matrix, then it falls in
one of the three cases of Lemma \ref{nihilpotent}. 
Assume that $A$ falls in case (2) of (3) of  Lemma \ref{nihilpotent}, we can compute explicitly the commutator of such 
matrices, but we find that all matrices that commute with $A$ are also nihilpotent. But, by hypothesis, there is a non
nihilpotent matrix in $\mathfrak g$ which commutes with $A$. Therefore, $A$ must fall in case (1) of Lemma \ref{nihilpotent}.
The space of matrices in that commute with $A$ is then easily computed and leas us to case 5.

In the third case, let us consider $G$ isomorphic to $\mathbb C\times \mathbb C$. Then, its Lie algebra $\mathfrak g$ is spanned
by two nihilpotent matrix. We have to split in two cases. If any matrix in $\mathfrak g$ falls in the case (3) of Lemma \ref{nihilpotent},
we can arrive easily to canonical form (5.b). If there is a matrix $A\in\mathfrak g$ that falls into case (3) of Lemma \ref{nihilpotent},
then $\mathfrak g$ is completely determined by $A$. Any other matrix in $\mathfrak g$ is in the commutator of $A$ which is a
Lie algebra of dimension $2$. The exponential of this Lie algebra lead us to canonical form (5.b). \hfill $\square$

\begin{cor}\label{Dim2}
Any maximal connected abelian subgroup of ${\rm Sp}(4,\mathbb C)$ is
of dimension $2$. 
\end{cor}

\section{Canonical Forms of Integrable Systems}

Let $H(t,x_1,x_2,y_1,y_2)\in \overline{\mathbb M(\Gamma)}[V]_2$ be an integrable quadratic homogeneous 
Hamiltonian of $2+\frac{1}{2}$ degrees of freedom. By theorem \ref{main} we know that the
notion of complete integrability and integrability in the non-autonomous sense are equivalent, 
so that we will just speak of an integrable non-autonomous Hamiltonian.   

We also know, by Theorem \ref{main} that its differential Galois group is a connected abelian
subgroup of ${\rm Sp}(4,\mathbb C)$. Proposition \ref{Group1} and Theorem \ref{Group2} give us
a complete list of all conjugacy classes of abelian subgroups of  ${\rm Sp}(4,\mathbb C)$. We can
then apply Lie-Kolchin reduction (Theorem \ref{KolchinReduction}) to $\vec X_{H}$ obtaining a 
canonical form for the Hamiltonian.

At this point we should remark that, when applying a time dependent 
change of frame to a non-autonomous Hamiltonian system, the Hamiltonian
function is then modified in the following form. If $B(t)\in {\rm Sp}(4,\overline{\mathbb M(\Gamma)})$ 
is a time dependent symplectic matrix that give us a change of frame, 
$$\left( \begin{array}{c} \xi_1 \\ \xi_2 \\ \eta_1 \\ \eta_2 \end{array}\right) =  
B(t)\left(\begin{array}{c}x_1 \\ x_2 \\ y_1 \\ y_2 \end{array}\right)$$
then, the new Hamiltonian function for our Hamiltonian systems is,
$$\bar H = H - (\xi_1, \xi_2 ,\eta_1, \eta_2)J\dot B(t)B^{-1}(t)\left(\begin{array}{c} \xi_1 \\ \xi_2 \\ \eta_1 \\ \eta_2 \end{array}\right), 
\quad J = \left(\begin{array}{cc} & I \\ -I &  \end{array}  \right),$$
as it follows from the change of frame formula for linear systems \ref{COF},
We can state directly the following result.

\begin{theorem}\label{Canonical}
Let $H(t,x_1,x_2,y_1,y_2)\in \overline{\mathbb M(\Gamma)}[V]_2$ be an integrable quadratic homogeneous 
Hamiltonian of $2+\frac{1}{2}$ degrees of freedom. Then, there exist a symplectic change of
frame,
$$\left( \begin{array}{c} \xi_1 \\ \xi_2 \\ \eta_1 \\ \eta_2 \end{array}\right) =  
B(t)\left(\begin{array}{c}x_1 \\ x_2 \\ y_1 \\ y_2 \end{array}\right)$$
with $B(t)\in {\rm Sp}(4,\overline{\mathbb M(\Gamma)})$ 
such that, for the transformed Hamiltonian $\bar H(\xi_1,\xi_2,\eta_1,\eta_2)$,
$$\bar H = H - (\xi_1, \xi_2 ,\eta_1, \eta_2)J\dot BB^{-1}\left(\begin{array}{c} \xi_1 \\ \xi_2 \\ \eta_1 \\ \eta_2 \end{array}\right), 
\quad J = \left(\begin{array}{cc} & I \\ -I &  \end{array}  \right)$$
belongs to one of the following categories:
\begin{center}
  \begin{tabular}{| c | c | c | c |}
    \hline
    Normal Form & Galois & Quadratic Invariants & Parameters \\ \hline \hline
     $0$ & $\{1\}$ & All &  \\ \hline
    $f(t)\left(\xi_1\eta_1 + \frac{p}{q}\xi_2\eta_2\right)$ & $\mathbb C^*$ & $\xi_1\eta_1$, $\xi_2\eta_2$ & 
    $f(t),\frac{p}{q}$\\ \hline
    $f(t)(\xi_1\eta_1 + \xi_2\eta_2)$ & $\mathbb C^*$ & 
    $\xi_1\eta_1$, $\xi_2\eta_2$, $\xi_1\eta_2-\xi_2\eta_1$ & $f(t)$ \\ \hline
    $f(t)\xi_1\eta_1$ & $\mathbb C^*$ & $\xi_1\eta_1$, $\xi_2^2$, $\eta_2^2$, $\xi_2\eta_2$ & $f(t)$\\ \hline
    $f(t)\frac{\eta_1^2}{2}$ & $\mathbb C$ & $\eta_1^2$, $\xi_2^2$, $\xi_2\eta_2$, $\eta_2^2$ & $f(t)$ \\ \hline
    $f(t)\frac{\eta_1^2+\eta_2^2}{2}$  & $\mathbb C$ & $\eta_1^2$, $\eta_2^2$ & f(t)\\ \hline
    $f(t)\left(\xi_2\eta_1 + \lambda\eta_1^2+\frac{\eta_2^2}{2}\right)$ & $\mathbb C$ & 
    $2\xi_2\eta_1 + \eta_2^2$, $\eta_1^2$ & $f(t)$, $\lambda$ \\ \hline 
    $f(t)\xi_1\eta_1 + g(t)\xi_2\eta_2$ & $(\mathbb C^*)^2$ & $\xi_1\eta_1$, $\xi_2\eta_2$ & $f(t)$, $g(t)$ \\ \hline
    $f(t)\frac{\eta_1^2}{2} + g(t)\xi_2\eta_2$ & $\mathbb C\times\mathbb C^*$ & $\eta_1^2$, $\xi_2\eta_2$ &
    $f(t)$, $g(t)$ \\ \hline
    $f(t)\frac{\eta_1^2}{2} + g(t)\eta_2^2$ & $\mathbb C^2$ & $\eta_1^2$, $\eta_2^2$ & $f(t)$, $g(t)$ \\ \hline
    $f(t)\eta_1\left(\xi_2 + g(t)\eta_1+\frac{\eta_2^2}{2}\right)$ & 
    $\mathbb C^2$ &  $2\eta_1\xi_2 + \eta_2^2$, $\eta_1^2$ & 
    $f(t)$, $g(t)$ \\ \hline
  \end{tabular}
\end{center}
Where $f(t)$ and $g(t)$ are arbitrary meromorphic functions, and $\lambda$ is an arbitrary constant,
and $p,q$ are coprime integers.

\end{theorem}

\section*{Acknowledgements}%
We want to thank to Primitivo Acosta-Hum\'anez, Guy Casale, Yuri Fedorov and Juan Jos\'e Morales-Ruiz 
for their helpful suggestions and comments, and to Felix Soriano and Kazuyuki Yagasaki 
for their support. This work was supported in part by MICINN-FEDER grant MTM2009-06973, 
CUR-DIUE grant 2009SGR859 and Universidad Sergio Arboleda.


{\noindent\sc David Bl\'azquez-Sanz\\
Instituto de Matem\'aticas y sus Aplicaciones\\
Universidad Sergio Arboleda \\
Bogot\'a, Colombia. \\
e-mail: } david@ima.usergioarboleda.edu.co

\bigskip

{\noindent\sc Sergio A. Carrillo Torres\\
Facultad de Ciencias\\
Departamento de Matem\'aticas\\
Universidad Nacional de Colombia\\
Bogot\'a, Colombia.  \\
e-mail: }  sacarrillot@unal.edu.co

\end{document}